\documentclass[twocolumn,superscriptaddress,floatfix,prb,amsmath,aps,showpacs]{revtex4}
 \usepackage{graphicx}
\usepackage{bm}

\begin{document}
\bibliographystyle{apsrev}

\def\nn{\nonumber}

\title{Quantum impurity spin in Majorana edge fermions} 
\date{\today}
\author{Ryuichi Shindou}
\affiliation{Condensed Matter Theory Laboratory, RIKEN,
2-1 Hirosawa, Wako, Saitama 351-0198, Japan}
\author{Akira Furusaki}
\affiliation{Condensed Matter Theory Laboratory, RIKEN,
2-1 Hirosawa, Wako, Saitama 351-0198, Japan} 
\author{Naoto Nagaosa} 
\affiliation{Department of Applied Physics, University of Tokyo, 
7-3-1 Hongo, Tokyo 113-8656, Japan} 
\affiliation{ 
%Cross-Correlated Materials Research Group (CMRG), 
%and Correlated Electron Research Group (CERG), 
CMRG and CERG, RIKEN-ASI, Wako 
351-0198, Japan}  
\begin{abstract}
We show that Majorana edge modes of two-dimensional spin-triplet
topological superconductors (superfluids) have Ising-like spin density
whose direction is determined by the $\bm{d}$-vector
characterizing the spin-triplet pairing symmetry.
Exchange coupling between an impurity spin ($S=\frac{1}{2}$) and
Majorana edge modes is thus Ising-type. 
Under external magnetic field perpendicular to the Ising axis, 
the system can be mapped to a two-level system with
Ohmic dissipation, which is equivalent to the anisotropic Kondo model. 
The magnetic response of the impurity spin
can serve as a local experimental probe for the 
order parameter.
\end{abstract}
\maketitle

%\paragraph{Introduction}
Majorana fermions are fermionic particles
that are their own antiparticle. 
Originally proposed long ago to describe neutrinos
in high energy physics, \cite{MF}
Majorana fermions have recently been a subject of intensive studies
in condensed matter physics. \cite{Wilczek,fk}
Their mixed nature of being particle and antiparticle implies that
Majorana fermions may emerge as elementary excitations
in superconductors and
superfluids where the number of particles is not well-defined.
Indeed they are theoretically predicted to appear
as gapless boundary excitations of
\textit{topological} superconductors/superfluids
with spin triplet pairing. \cite{srfl,kitaev}
Candidates of such topological materials include 
superfluid phases of $^3$He, \cite{v}
the superconducting states of Sr$_2$RuO$_4$, \cite{mi} 
%(TMTSF)$_2$X salt~\cite{TMTSF},
and possibly some 
non-centrosymmetric superconductors. \cite{nonC}
However, being ``real-part" of ordinary (complex) fermions,
Majorana fermions are charge neutral and 
only weakly interacting with other particles;
they are hard to detect.
To probe and control Majorana fermions is therefore
a great challenge.

In this paper we study
a quantum impurity spin coupled with 
Majorana edge modes with spin degree of freedom.
We first argue that a distinct signature of Majorana edge 
fermions of a generic two-dimensional (2D) spin-triplet 
topological superconductor (superfluid) is Ising character of their
spin density. \cite{sr,cz}
To probe this Ising spin, we consider a spin-$\frac12$
magnetic impurity coupled to the Majorana edge modes.
We show that the impurity spin has strongly anisotropic and
singular magnetic response which is due to
{\it quantum dissipation} from
the Majorana edge modes.
We propose that electron spin resonance (ESR)
can serve as 
a novel local probe for the Majorana fermions 
in spin-triplet topological superconductors and superfluids.

%\paragraph{Majorana Ising spin}
Let us take a closer look at the Majorana edge modes
of 2D spin-triplet topological superconductors.
We first consider the BdG Hamiltonian of
a prototypical topological superconductor,
a \textit{spinless} chiral $p$-wave superconductor
with $p_x\pm ip_y$ symmetry, \cite{rgi}
\begin{equation}
\mathcal{H}^{\vartheta}_{\pm} =
\left(\begin{array}{cc}
\displaystyle
-\frac{\hbar^2}{2m}\partial^2 -\mu &
\displaystyle
\frac{e^{i\vartheta}}{2ik_F} \{ \Delta(\bm{r}), \partial_\pm \} \\
\displaystyle
\frac{e^{-i\vartheta}}{2ik_F} \{ \Delta(\bm{r}),\partial_\mp \} & 
\displaystyle
\frac{\hbar^2}{2m}\partial^2 + \mu \\
\end{array}\right).
\label{bdg1}
\end{equation}
Here
$\bm{r}=(x,y)$,
$\partial_{\pm}=\partial_x \pm i \partial_y$,
$\partial^2=\partial_+\partial_-$,
$k_F$ the Fermi wave number, $\mu=\hbar^2k_F^2/2m$,
and $\vartheta$ is a U(1) phase.
In the bulk superconductor with a spatially uniform pair
potential $\Delta(\bm{r})=\Delta$,
quasiparticle spectrum is
fully gapped.
When the superconductor has a boundary,
the BdG Hamiltonian has a gapless edge mode 
with linear energy dispersion 
$E_k = vk = \Delta k/k_F$ ($|k|<k_F$).
For a straight boundary defined by
$X\equiv  x \cos\phi + y \sin\phi=0$,
we solve the BdG Hamiltonian on the half plane $X<0$
[$\Delta(\bm{r}) = \Delta \Theta(-X)$],
assuming fixed boundary conditions at $X=0$.
We then obtain the edge mode's wave function
\begin{eqnarray}
\left(\begin{array}{c}
u_{k}(\bm{r}) \\ 
v_{k}(\bm{r}) \\
\end{array}\right) 
= e^{\pm i k Y} w_{k}(X) 
\left(\begin{array}{c} 
e^{i(2\vartheta \pm 2\phi+\pi)/4} \\
e^{-i(2\vartheta \pm 2\phi +\pi)/4} \\
\end{array}\right).
\label{ab}   
\end{eqnarray}
Here $Y$ is the coordinate 
along the edge,  
$Y\equiv -x \sin\phi + y \cos\phi $, 
and $w_k(X)$ is the normalized 
real-valued wave function localized at the edge.
The particle-hole symmetry, 
$\mathcal{P}\mathcal{H}^\vartheta_\pm \mathcal{P}^\dagger
=-\mathcal{H}^\vartheta_\pm$ with $\mathcal{P}=\sigma_1 K$, 
guarantees that
$(v^{*}_{k},u^{*}_{k})=(u_{-k},v_{-k})$ 
is also an eigenmode with energy $-vk$,
where $K$ is complex conjugation and
$\sigma_j$ is the $j$th component of the Pauli matrices.
The mode expansion of the Nambu field 
$(\psi,\psi^{\dagger})^{t}$ for $|E|<\Delta$ is then given by
\begin{eqnarray}
&&\hspace{-0.6cm}
\left(\begin{array}{c} 
\psi(\bm{r}) \\
\psi^{\dagger}(\bm{r}) \\
\end{array}\right)\! 
= \! \int^{k_F}_0\! dk
\bigg[\hat{\gamma}_k 
\!\left(\begin{array}{c} 
u_k (\bm{r}) \\ 
v_k (\bm{r}) \\
\end{array}\right)\!
+ \hat{\gamma}^{\dagger}_k 
\left(\begin{array}{c} 
v^{*}_k (\bm{r}) \\ 
u^{*}_k (\bm{r}) \\
\end{array}\right)\!
\bigg],
\label{mode}
\end{eqnarray} 
which leads to the following condition of Majorana type 
\begin{eqnarray}
\psi(\bm{r}) = ie^{i\vartheta \pm i\phi} 
\psi^{\dagger}(\bm{r}). \label{majo}  
\end{eqnarray}

There are two distinct types of 2D
\textit{spin-triplet} topological superconductors:  
\cite{srfl,kitaev} (a) {\it chiral} type \cite{v} 
and (b) {\it helical} type. \cite{srfl,Roy,Qi} 
They can be obtained, for example, by combining
two copies of the spinless chiral
$p$-wave superconductors of
same or opposite chiralities;
their BdG Hamiltonians are given by
${\cal H}^{\theta_{\uparrow}}_{+} 
\oplus {\cal H}^{\theta_{\downarrow}}_{+}$ 
and 
${\cal H}^{\theta_{\uparrow}}_{+}\oplus 
{\cal H}^{\theta_{\downarrow}}_{-}$, respectively.
The two types of 2D spin-triplet topological superconductors
are characterized by the order parameter 
$\hat{\Delta}_{\bm{k}} = i{\bm d}_{\bm{k}}\cdot{\bm \sigma}\sigma_2$
with the ${\bm d}$-vector,~\cite{ajl} 
\begin{equation}
{\bm d}_{\bm{k}}
= \left\{ 
\begin{array}{ll} 
(\hat{\bm x} s_{\theta} + \hat{\bm y} c_{\theta}) (k_x + i k_y),
& \mbox{(chiral)},\\
(\hat{\bm x} s_{\theta} + \hat{\bm y} c_{\theta}) k_{x}
 + (\hat{\bm x} c_{\theta} - \hat{\bm y} s_{\theta}) k_y,
& \mbox{(helical)}.
\end{array}\right.
\label{bw}
\end{equation}
Here $(c_{\theta},s_{\theta})\equiv (\cos\theta,\sin\theta)$,
$\theta=\frac{1}{2}(\theta_{\uparrow}-\theta_{\downarrow})$,
and $\hat{\bm x}$ and $\hat{\bm y}$ are unit vectors in the spin space.
Notice that the direction of 
$\bm{d}_{\bm k}$ is independent of ${\bm k}$
in the chiral case, while it constitutes a 
coplanar spin texture in the ${\bm k}$-space
in the helical case.

The two types of 2D spin-triplet topological superconductors
support gapless Majorana edge modes,
each spin component of which
obeys Eq.~(\ref{majo}):
(a) $\psi_{\sigma} ({\bm r})= 
ie^{i\theta_{\sigma} + i\phi} \psi^{\dagger}_{\sigma} ({\bm r})$ 
($\sigma=\uparrow$ and $\downarrow$) for the chiral case, 
and (b) $\psi_{\uparrow}({\bm r})= 
ie^{i\theta_{\uparrow}+ i\phi} \psi^{\dagger}_{\uparrow}({\bm r})$
and $\psi_{\downarrow}({\bm r})=
ie^{i\theta_{\downarrow} - i\phi} \psi^{\dagger}_{\downarrow}({\bm r})$ 
for the helical case. 
These Majorana conditions lead to the operator identities 
for the edge mode's spin density, 
\begin{subequations}
\begin{eqnarray} 
2\hat{s}_z (\bm{r}) \!\!&=&\!\!
 \psi^{\dagger}_{\uparrow}\psi_{\uparrow}^{} 
- \psi^{\dagger}_{\downarrow}\psi_{\downarrow}^{} = 0 ,  \\  
\hat{s}_{+}(\bm{r}) 
\!\!&=&\!\! \psi^{\dagger}_{\uparrow}\psi_{\downarrow}
= \left\{ 
\begin{array}{ll} 
-e^{-2i\theta} \hat{s}_{-}(\bm{r}) & 
({\rm chiral}), \\
-e^{-2i(\theta + \phi)} \hat{s}_{-}(\bm{r}) & 
({\rm helical}). \\ 
\end{array} \right. \label{Ising-spin-b}  
\end{eqnarray}
\end{subequations}
These equations imply that the spin density is always Ising-like:  
$(\hat{s}_x,\hat{s}_y)\propto(s_\theta,c_\theta)$
for the chiral case and  
$(\hat{s}_x,\hat{s}_y)\propto(s_{\theta+\phi},c_{\theta+\phi})$
for the helical case. 
Comparing Eq.~(\ref{bw}) with Eq.~(\ref{Ising-spin-b}), 
we can see that this Ising direction is 
dictated by the ${\bm d}$-vector in the bulk as,
\begin{eqnarray} 
\hat{\bm s}(\bm{r}) \propto  \left\{ 
\begin{array}{ll} 
{\bm d}_{\bm{k}} & ({\rm chiral}), \\
{\bm d}_{\hat{\bm{X}}} &
 ({\rm helical}), \\ 
\end{array} \right. \label{Ising-spin} 
\end{eqnarray}
where
$\hat{\bm{X}}=(\cos\phi,\sin\phi)$ is a vector normal to the boundary.
In the helical case,
$\hat{\bm s}(\bm{r})$ depends
on the direction of the boundary,
as the rotation in the real space is transcribed
into that in the spin space. Otherwise, it 
does not depend on the shape of 
the boundary. 

The Ising-like spin density (\ref{Ising-spin}) is
a hallmark of the Majorana edge modes.
Its strong anisotropy reflects the spin-triplet pairing
symmetry in the bulk topological superconducting order.
In the following we will study quantum dynamics of
a magnetic impurity (probe spin) interacting
with this Ising spin density.

%\paragraph{The model}
The Hamiltonian for the
coupling between the Majorana Ising spin density $\hat{\bm{s}}(\bm{r})$
and a spin-$\frac12$ probe at $\bm{r}=0$,
$\hat{\bm{S}}=(\hat{S}_x,\hat{S}_y,\hat{S}_z)$,
is given by
\begin{eqnarray}
{\cal H}_{\rm ex} = i J \hat{S}_z
\tilde{\psi}_{\uparrow}(0)\tilde{\psi}_{\downarrow}(0). \label{sd}
\end{eqnarray}
Here the probe spin's $S_z$ direction is taken to be
parallel to $\hat{\bm{s}}(0)$, and
$\tilde{\psi}_\sigma(Y)$ is a one-dimensional (1D) Majorana field
satisfying
$\tilde{\psi}^{\dagger}_{\sigma} = \tilde{\psi}_{\sigma}$
and
$\{\tilde{\psi}_\sigma(Y),\tilde{\psi}^\dagger_{\sigma'}(Y')\}
=\delta_{\sigma,\sigma'}\delta(Y-Y')$.
The 1D Majorana field is obtained from $\psi_\sigma(\bm{r})$
by appropriate U(1) gauge transformation and dropping 
unimportant $X$-dependence [i.e., $w_k(X)\to1$].
%In the case of  topological {\it superconductors}, 
The Kondo coupling in Eq.\ (\ref{sd}) 
can be obtained from the Anderson 
impurity model,
\[
{\cal H}_{\rm imp} = 
\sum_{\sigma=\uparrow,\downarrow} \!
\left\{
\epsilon_d n_{d,\sigma}
+ 
%\sum_{\sigma=\uparrow,\downarrow}
t\left[d^{\dagger}_{\sigma} 
\tilde{\psi}_{\sigma}(0) + {\rm h.c.}\right]\right\} \!
+ U n_{d,\uparrow} n_{d,\downarrow}
\]
with 
$n_{d,\sigma}\equiv d^{\dagger}_{\sigma}d_{\sigma}$,
by the standard procedure.
This yields
$\hat{S}_z \equiv - \frac{1}{2}  \!\  
d^{\dagger}_{\alpha} [\sigma_y]_{\alpha\beta} 
d_{\beta}$ 
and
$J = 2 t^2U/[(U+\epsilon_d)|\epsilon_d|] \, (>0)$.
The kinetic energy of 
the Majorana edge modes reads
\begin{equation}
{\cal H}_{\rm kin} = iv
\int^{\infty}_{-\infty} \!
dY \left(
\tilde{\psi}^{\dagger}_{\uparrow} \partial_Y 
\tilde{\psi}_{\uparrow} 
\pm \tilde{\psi}^{\dagger}_{\downarrow} \partial_Y
\tilde{\psi}_{\downarrow} \right),
\end{equation}
where the $+/-$ signs 
in the integrand
are for the chiral/helical superconductors,
respectively. 
The ground state of
$\mathcal{H}_{\rm kin}+\mathcal{H}_{\rm ex}$ is
doubly degenerate, $S_z=\pm\frac12$. 

We will make full use of the knowledge from
earlier studies on dissipative two-state systems and Kondo
effect, \cite{l-w,weiss,ayh}
to show that the impurity spin,
when subjected to external magnetic fields,
exhibits anisotropic dissipative quantum dynamics
due to the ``background'' Majorana edge modes.

\begin{figure}
    \includegraphics[width=72mm]{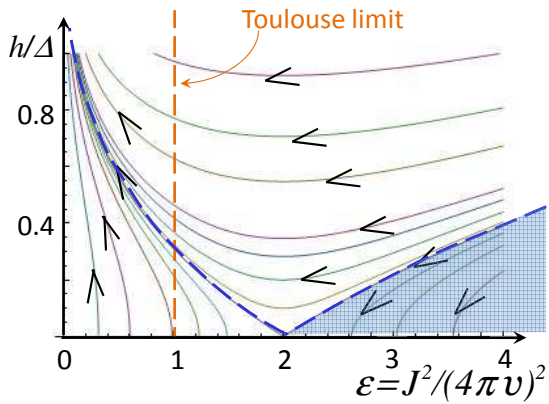}
\caption{(color online) 
Renormalization group flows.}
\label{fig:RGflow}
\end{figure}

Consider first that a magnetic field is applied 
{\it perpendicular} to the Majorana 
Ising spin, say along the $x$-direction;
the Hamiltonian reads 
\begin{eqnarray}
{\cal H} = {\cal H}_{\rm kin} + {\cal H}_{\rm ex} 
+ h \hat{S}_x. 
\label{ham}
\end{eqnarray}
Interestingly, the system undergoes a quantum 
phase transition,  depending the exchange coupling $J$ 
(Fig.~1). 
To see this, let us map the Hamiltonian 
onto the Ohmic dissipative two-state 
system,~\cite{l-w,weiss} 
\begin{eqnarray}
{\cal H}_{\rm ts} = v\int^{\infty}_{-\infty} 
(\partial_y \Phi)^2 dy 
+ \frac{J}{\sqrt{2\pi}} \hat{S}_z (\partial_y \Phi)|_{y=0}
+ h \hat{S}_x,
\label{tss} 
\end{eqnarray}
where the bosonic field $\Phi(x)$ obeys
$[\Phi(x),\partial_y \Phi(y)] = i\delta(x-y)$. 
We have combined the two species of chiral Majorana fields, 
$\{\tilde{\psi}_{\uparrow},\tilde{\psi}_{\downarrow}\}$, 
into a single spinless chiral fermion,
$\Psi(y)=[\tilde{\psi}_{\uparrow}(y)
+i\tilde{\psi}_{\downarrow}(\pm y)]/\sqrt{2}$,
where the $+/-$ sign refers to the chiral/helical superconductors,
respectively.
The Ising spin density at $\bm{r}=0$ is then reduced to
the fermion number density $\Psi^\dagger(0)\Psi(0)$.
The complex fermionic field is readily 
bosonized in terms of the phase operator, 
$\Psi(y) \propto \exp[i\sqrt{2\pi}\Phi(y)]$,
which transforms Eq.~(\ref{ham}) into Eq.~(\ref{tss}).
The Ising exchange coupling $J$ and the transverse field $h$
control the coupling to the Ohmic bath and tunneling between
the two states ($S_z=\frac12$, $-\frac12$), respectively.
The dissipative two-state system is related 
to the anisotropic Kondo model,~\cite{l-w,weiss} 
\begin{eqnarray}
{\cal H}_{\rm K} \!\!&=&\!\!
-2iv\int dy\left[
\Psi^\dagger_\uparrow(y)\partial_y\Psi_\uparrow(y)
+\Psi^\dagger_\downarrow(y)\partial_y\Psi_\downarrow(y)
\right]
\nonumber\\
&&\!\!{}
+J_\perp\!\left[
\hat{S}_+\Psi^\dagger_\downarrow(0)\Psi_\uparrow(0)
+\hat{S}_-\Psi^\dagger_\uparrow(0)\Psi_\downarrow(0)\right]
\nonumber\\
&&\!\!{}
+J_z\hat{S}_z\!\left[
\Psi^\dagger_\uparrow(0)\Psi_\uparrow(0)
-\Psi^\dagger_\downarrow(0)\Psi_\downarrow(0)
\right],
\label{akm}
\end{eqnarray}
where $\Psi_{\uparrow,\downarrow}$ are
complex spinful fermionic fields
and $\hat{S}_\pm=\hat{S}_x \pm i \hat{S}_y$.
The coupling constants are related by 
%$h\tau=(J_{\perp}/4\pi v)$ and  
%$(J/2 \pi v) = \sqrt{2}\left((J_z/2\pi v)-2\right)$,   
\begin{eqnarray}
h\tau=\frac{J_{\perp}}{4\pi v}, \ \  \ 
\frac{J}{2 \pi v} = \sqrt{2}\left(\frac{J_z}{2\pi v}-2\right),   
\end{eqnarray} 
where
$\tau$ is inverse of an
ultraviolet cutoff (band width) of 
the massless Majorana edge modes, i.e.,
$\tau^{-1} = \Delta$. 
The transverse $(J_\perp)$ and longitudinal ($J_z$) Kondo exchange
couplings are related 
to the transverse field $h$ and the Ising exchange
coupling $J$, respectively.
With the help of these mappings, 
the renormalization group (RG) 
equations for $\eta\equiv h \tau$ 
and $\epsilon \equiv (J/4\pi v)^2$ 
can be readily obtained,~\cite{ayh} 
\begin{eqnarray}
\frac{d\epsilon}{d\ln \tau} = -4\epsilon \eta^2, 
\qquad
\frac{d\eta}{d\ln \tau} = \frac{1}{2}(-\epsilon+2)\eta.  
\label{rg}
\end{eqnarray}
These scaling equations lead to the RG flow diagram
in Fig.~\ref{fig:RGflow}, which has
a fixed point at $(J,h)=(0,\infty)$ and 
a line of fixed points at $h=0$ for
$|J|>4\sqrt{2}\pi v \equiv J_c$ ($\epsilon>2$).
When $|J|<J_c$, the transverse field $h$ is a relevant perturbation,
and the system flows toward the high-field fixed point.
In this case the ground 
state is a quantum mechanical superposition of 
$S_z=+\frac{1}{2}$ and $-\frac{1}{2}$ states,
corresponding to the antiferromagnetic Kondo singlet.
When $|J|>J_c$, on the other hand,
the orthogonality catastrophe of Majorana edge excitations
suppresses the tunneling, thereby making a small transverse field
$h$ irrelevant.
For small fields $h<h_c$,
where
\begin{equation}
h_c 
= \frac{\Delta}{\sqrt{2}}\Big(\frac{|J|}{4\pi v} - \sqrt{2}\Big), 
\label{hc} 
\end{equation}
the RG flows end at the zero-field fixed line;
this corresponds to the ferromagnetic Kondo regime, $J_z\le-|J_\perp|$.
For large enough fields $h>h_c$, the system is eventually 
renormalized to the strong-coupling regime
(see Fig.~\ref{fig:RGflow}).
The crossover from weak- to strong-field regime occurs
around the Kondo temperature $T_K$ given as,~\cite{ayh,l-w}   
\begin{eqnarray}
T_K = \left\{ \begin{array}{ll}
\Delta \exp(-b/\sqrt{h-h_c}\,),
& |J|>J_c\ \cap \ h>h_c, \\
h \big(h/\Delta\big)^{\epsilon/(2-\epsilon)}, 
& |J| < J_c, \\ 
\end{array} \right. \label{k1}
\end{eqnarray}
where $b=\pi\Delta/\sqrt{8h_c}$.

%\paragraph{Longitudinal responses}
Let us assume that the magnetic field has a component
in the direction of the Majorana Ising spin.
This adds a weak Zeeman term $\mathcal{H}_z=-h_z\hat{S}_z$
to the Hamiltonian.  From the mapping to the anisotropic 
Kondo model,~\cite{ayh,l-w,weiss,fm} 
we see that the response of the probe spin $\hat{\bm{S}}$ to a weak
longitudinal field $h_z\hat{\bm z}$
should be singular  
at the critical field $h=h_c$ for $|J|>J_c$.
In the ferromagnetic Kondo regime where $h\le h_c$ and $|J|>J_c$,
the ground state is doubly degenerate, and
the magnetization $M_z=\langle\hat{S}_z\rangle$ shows a jump
across $h_z=0$ at zero temperature. \cite{fm}
As $h$ is increased, the magnetization jump is reduced but
remains finite at $h=h_c$. \cite{fm} 
At finite temperature $T$, the static longitudinal susceptibility
$\chi_{zz}=\partial M_z/\partial h_z|_{h_z=0}$ shows a 
Curie behavior, $\chi_{zz}\propto1/T$, 
as long as $h\le h_c$.
Above the critical field $h_c$, the ground state is a singlet.
In this case, the longitudinal susceptibility saturates
below the Kondo temperature,
$\chi_{zz}\approx1/T_K$ at $T<T_K$.
The saturated value diverges at $h= h_c$.

Ac longitudinal field, $h_z \hat{\bm z} e^{-i\omega t}$,
induces transitions between the levels split by
the tunneling term $h\hat{S}_x$.
To see this, let us consider the dynamical 
susceptibility 
%\begin{equation}
$\chi_{zz}(\omega) 
= i \int^{\infty}_{0} dt  
  e^{i(\omega+i0)t}\langle [\hat{S}_z(t),\hat{S}_z(0)]\rangle$, 
%\end{equation},  
which is a Fourier transform of the time-evolution of
the magnetization,
$P(t)=\langle \hat{S}_z(t)\rangle$, studied in the context
of the dissipative two-state system. \cite{l-w,weiss} 
The transition spectrum is obtained from the imaginary part
of $\chi_{zz}(\omega)$
and has a universal scaling form which depends on $\epsilon$, 
$S_{zz}(\omega) :=
\mathrm{Im}\chi_{zz}(\omega)/\omega   
= f_{\epsilon} (\omega / T_K)$. 
Its qualitative feature is well understood (for small $h/\Delta$), 
\cite{l-w,weiss,cklss} as we briefly summarize below.

\begin{figure}
    \includegraphics[width=60mm]{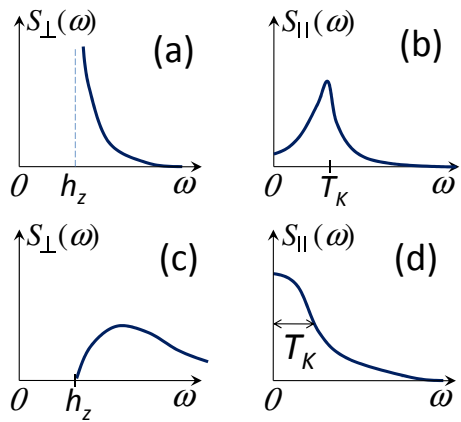}
\caption{(color online) Transition (absorption) spectra.  
(a) $S_\perp(\omega)=\mathrm{Im}\,\chi_{xx}(\omega)/\omega$ 
under longitudinal dc field $h_z$ and weak dissipation $\epsilon<1$.
(b) $S_\parallel(\omega)=\mathrm{Im}\,\chi_{zz}(\omega)/\omega$
under transverse dc field $h$ and weak dissipation $\epsilon<\frac23$.
(c) $S_\perp(\omega)$ for $\epsilon\ge1$, and
(d) $S_\parallel(\omega)$ for $\epsilon\ge\frac23$
($T_K\equiv0$ for $\epsilon>2$).
}
\label{fig:xray}
\end{figure}

In the weakly dissipative regime $\epsilon<\frac{2}{3}$, 
$S_{zz}(\omega)$ has a peak at
$\omega \approx T_{K}$, which signifies
coherent transitions between the bonding and anti-bonding states
split by the renormalized transverse field (Fig.~\ref{fig:xray}b).
Stronger dissipation destroys the coherent tunneling;
when $\epsilon>\frac{2}{3}$, 
the coherence is lost and $S_{zz}(\omega)$
shows only a diffusive peak at $\omega=0$ 
(Fig.~\ref{fig:xray}d). 
The half-value width of this peak
is on the order of $T_K$. \cite{cklss}   
As the dissipation is further increased toward
the ferromagnetic Kondo region, the diffusive peak gets sharper,  
as $T_K \rightarrow 0$ at $\epsilon=2$.

In the ferromagnetic Kondo region,
the spectral function at $T=0$
has a $\delta$-function peak at $\omega=0$.~\cite{l-w,weiss}   
At finite temperature, or
in the presence of a finite dc ``bias'' field $h_z$,
the width of the diffusive peak at $\omega=0$
is finite, as $T$ and $h_z$ play a role of low-energy
cutoff. \cite{l-w,weiss}  
The half-value width $\omega_0$ can be obtained by
2nd order perturbation in the transverse field $h$,
yielding
$\omega_{0} \propto
(h^2/\Delta)\!\ 
(T/\Delta)^{\epsilon-1}$ or
$(h^2/\Delta)(h_z/\Delta)^{\epsilon-1}$.  
A diffusive peak with the width of the same $T$ and 
$h_z$-dependences also appears in the antiferromagnetic 
Kondo regime,
when $T > T_K$ or $h_z > T_K$.  

%\paragraph{Transverse responses} 
Next, we discuss response of the probe spin $\bm{S}$
to the transverse field $h$ (while we set $h_z=0$),
such as the magnetization $M_x=\langle \hat{S}_x\rangle$
and the susceptibility $\chi_{xx}=\partial M_x/\partial h$.
This response is unique to our system and has not been
considered in the related two models (\ref{tss}) and (\ref{akm}),
as the Zeeman term $h\hat{S}_x$ corresponds to the tunneling
and the Kondo exchange interaction, respectively.

At the critical field $h=h_c$ separating the ferromagnetic
and antiferromagnetic Kondo phases,
the transverse magnetization $M_x$ shows only weak
(essential) singularity,
$M_x(h)-M_x(h_c) \propto \exp(- b/\sqrt{|h| - h_c} )$.
This singularity is like the specific heat anomaly at
Kosterlitz-Thouless transition
and is too weak to be observed.
Thus $M_x$ increases monotonically as a function of 
$h$ without any hint of anomaly at $h=h_c$.

In the limit of small transverse field $h\ll\Delta$,
the transverse magnetization $M_x$ increases linearly
with $h$, when dissipation is strong, $\epsilon>1$.
The susceptibility $\chi_{xx}$ is diverging
as $(\epsilon-1)^{-1}$
for $\epsilon\to1$, 
and at the point $\epsilon=1$,
the linear $h$ dependence acquires a logarithmic correction,
$M_x %= -2 h\Delta^{-1} {\rm Ei}\big(-\pi h^2\Delta^{-2}\big) 
= - 4 h\Delta^{-1} \ln(h/\Delta) + \cdots$.
This result can be obtained from the exact ground state energy
in the Toulouse limit. \cite{weiss} 
In the weakly dissipative case $0 < \epsilon <1$, 
the $h$ dependence becomes sublinear,
$M_x \propto \left(h/\Delta\right)^{\epsilon/(2-\epsilon)}$. 
This indicates that the linear susceptibility diverges 
in the low-$T$ limit.
However, this singularity is weaker than that 
of the Curie behavior and is given by
\begin{eqnarray}
\left.\chi_{xx}\right|_{h=h_z=0} =
\frac{\Gamma\big(\frac{1-\epsilon}{2}\big)}
{\sqrt{\pi}\,\Gamma\big(\frac{2-\epsilon}{2}\big)}
\Big[
\sin\!\Big(\frac{\pi T}{\Delta}\Big)
\Big]^{\epsilon}  
\frac{1}{T}
\propto T^{-1+\epsilon}. \label{curies}
\end{eqnarray}
In the presence of dc longitudinal field $h_z\hat{\bm z}$, 
a small ac transverse field $h \hat{\bm x} 
e^{-i\omega t}$ induces 
transitions between the $S_z=\pm\frac12$ states.
The transition spectrum has a divergent edge singularity, 
~\cite{gdm} 
\begin{eqnarray}
\left.{\rm Im}\chi_{xx}(\omega)\right|_{T=h=0}
= \frac{c\!\ \pi}{\Gamma(\epsilon)}\frac{e^{-c'|\omega-h_z|/\Delta}}
{\Delta^{\epsilon}|\omega-h_z|^{1-\epsilon}} 
\Theta(\omega-h_z),
\label{dyn-tr}
\end{eqnarray}
for positive $\omega$, 
where $c$ and $c^{\prime}$ are some positive constants. 
Equation (\ref{dyn-tr}) is
also valid for $\epsilon\ge 1$.

%\paragraph{Probe for the order parameter} 
Based on the results discussed so far, we 
propose that ESR measurements of
the impurity spin may be used to identify
the direction of the Majorana Ising spin.
A dc field smaller than the (lower) critical 
field $H_{c(1)}$ 
will cause precession of only those probe spins located 
around the boundary (edge). 
An additional ac field (perpendicular to the dc field)
will then induce resonant transitions.
Figure \ref{fig:xray} shows qualitative picture of
the absorption spectra for two complementary experimental geometries,
i.e., the dc field applied parallel (a,c) and
perpendicular (b,d) to the Majorana Ising spin.
The upper two panels (a,b) assume weak dissipation.
The spectra show the strongest edge singularity at the 
Larmor frequency $h_z$, when the dc field is
parallel to the Majorana Ising spin (a).
When the spin precession is driven
by the transverse field $h$ (b),
the singularity is replaced by a resonance peak at
$\omega=T_K = h (h/\Delta)^{\epsilon/(2-\epsilon)} \ll h$.
The clear difference between the spectra measured in the two 
experimental geometries persists at stronger dissipation
(compare c and d).
We thus expect that the Ising direction can be identified in principle
by measuring the absorption spectra for various directions of 
the dc field. Incidentally, anisotropy in the static susceptibility
may also serve as a good indicator; it shows Curie-law along 
the Ising direction ($\chi_{zz}$), while it remains constant 
for $\epsilon < 1$ or  
diverges as $T^{\epsilon-1}$ for $\epsilon > 1$ 
in the perpendicular direction ($\chi_{xx} \equiv \chi_{yy}$). 

In summary, 
we proposed to probe the massless Majorana edge 
modes of spin-triplet topological superconductors and 
superfluids,  by introducing a quantum impurity spin to 
their boundary. %of the topological superconductors.
Anisotropic magnetic responses of the probe spin 
directly reflect the Majorana nature of the edge excitations.

This work was supported 
by Grants-in-Aid for Scientific Research 
(Grants No.~17071007, 17071005, 19048008, 19048015,
21244053, and 21540332) from the MEXT and the JSPS, Japan.
N.N. is supported also by Funding Program for World-Leading 
Innovative R $\&$ D on Science and Technology (FIRST Program).

\end{document}